\documentclass{PoS}
\usepackage[caption=false]{subfig}

\title{Evolving Antennas for Ultra-High Energy Neutrino Detection}

\ShortTitle{Evolving Antennas for Ultra-High Energy Neutrino Detection}

\author{\speaker{Julie A. Rolla} for the GENETIS Collaboration\footnote{for collaboration list see PoS(ICRC2019)1177} 
        \\The Ohio State University\\
        E-mail: \email{rolla.3@osu.edu} }

\abstract{Evolutionary algorithms borrow from biology the concepts of mutation and selection in order to evolve optimized solutions to known problems. The GENETIS collaboration is developing genetic algorithms for designing antennas that are more sensitive to ultra-high energy neutrino-induced radio pulses than current designs. There are three aspects of this investigation. The first is to evolve simple wire antennas to test the concept and different algorithms. Second, optimized antenna response patterns are evolved for a given array geometry. Finally, antennas themselves are evolved using neutrino sensitivity as a measure of fitness. This is achieved by integrating the XFdtd finite-difference time-domain modeling program with simulations of neutrino experiments. }

\FullConference{36th International Cosmic Ray Conference -ICRC2019-\\
		July 24th - August 1st, 2019\\
		Madison, WI, U.S.A.}
\tableofcontents
\begin{document}

\section{Introduction}\label{S:Intro}
Neutrino astronomy is a field of particle astrophysics that seeks to investigate distant phenomena through the detection of neutrinos. As weakly interacting particles, neutrinos are able to travel cosmic distances directly from their source, thus providing a unique means for exploring particle sources in the distant universe \cite{Halzen2006, Gorham2019}. The field of ultra-high energy (UHE) neutrino astronomy probes the most energetic events at above $10^{18}$~eV. Due to the extremely low neutrino cross-section, UHE neutrino detection experiments rely on massive detector volumes, such as the Antarctic ice, to detect when a neutrino directly strikes an atom. The collision of a neutrino within the ice produces Askaryan radiation. Askaryan radiation is of particular interest, as it produces radio signals with attenuation lengths in pure ice on the order of 1~km \cite{Connolly2016}. 

The GENETIS (Genetically Evolving NEuTrIno TeleScopes) collaboration focuses on using genetic algorithms (GAs) to improve the sensitivities of Askaryan radiation-based detectors, such as those used by ANITA, ARA, and ARIANNA. These experiments utilize various types of antennas to detect the neutrino induced radio waves \cite{Gorham2019, Allison2016,  Anker2019}. GAs are computational techniques that mimic the mechanisms of biological evolution by creating populations of potential solutions to a defined problem and then evolving the population over multiple generations to find optimized solutions \cite{Davis1991, Kumar2010}. Our research seeks to improve experimental sensitivities with consideration to current constraints in geometry of antenna devices deployed in the ice, and the signal characteristics of neutrino generated Askaryan radiation. 

The proceedings describe some background information regarding GAs, as well as three main GA procedures for evolving antennas: (1) evolving simplified paperclip antennas, (2) the Antenna Response Evolution Algorithm (AREA), and (3) the Physical Antenna Evolution Algorithm (PAEA). Paperclip evolution is working towards evolving highly directional antennas. AREA is focused on evolving gain patterns that are better suited for neutrino detection. In PAEA, the physical properties of the antenna are evolved for improved neutrino sensitivities. In the future, these procedures could be combined, so that the evolution of the physical antennas produce a desired antenna response which itself is derived from an evolutionary algorithm.


\section{Genetic Algorithms}
GAs are a type of evolutionary computation used to discover one or more sets of values that provide a best-fit to a large parameter space \cite{Davis1991}. The resulting, multiple generation convergence can evolve solutions for problems that would have otherwise been difficult or not possible through more traditional techniques \cite{Davis1991, Zebulum2018}.

Each generation consists of a population of individuals, each of which is tested for its output against a predefined set of goals. To quantify the performance of an individual, a fitness function is created to generate a fitness score by comparing the genes of an individual to the optimal or desired goals. 
Genes hold characteristics of individuals. Each generation of individuals has the potential, but is not guaranteed, to improve upon the prior \cite{Shukla2015}.

\begin{figure}
    \centering
    \includegraphics[width=0.3\linewidth]{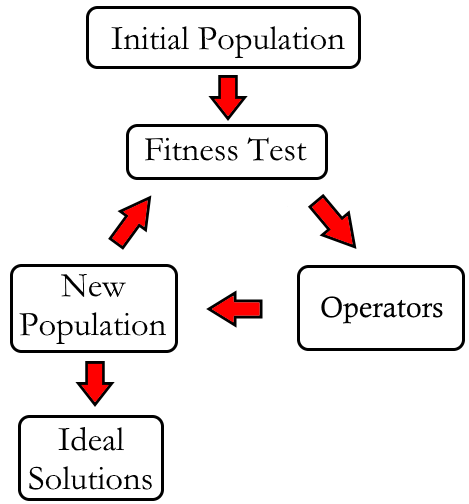}
    \caption{General GA evolution procedure.}
    \label{fig:GAFlowChart}
\end{figure}

The GA workflow is presented in Figure \ref{fig:GAFlowChart}. First, a population of individuals is generated, each with randomly generated genes. Each individual is tested through the fitness function to generate a fitness score. A selection method is implemented to decide which individuals, called parents, will be used in creating the next generation. An operator then uses the parents' genetic code to create offspring individuals for the next generation. The fitness test is applied to each individual, generating the new fitness scores. This process is iterated until a specified fitness score threshold is surpassed by one or more individuals, or until a specified number of generations have passed.

GAs often combine different selection methods and operators. The two most common selection methods are roulette and tournament selection. In roulette selection, the probability of an individual being selected is proportional to the individual's fitness score. In tournament selection, individuals are randomly placed into groups (tournaments) and compared by fitness scores. The individual(s) with the highest fitness score in each tournament are then selected \cite{Methods}. Additionally, two primary operators exist: crossover and mutation. In crossover, individuals swap genes to form offspring. In mutation, genetic diversity is introduced to avoid convergence to an incorrect solution by randomly altering genes \cite{Mutation}.


\section{Paperclip Antenna Evolution} 
GAs can be used to design antennas with desired gain patterns. An initial investigation of this process was conducted on a simple, segmented antenna design, modeled on the antenna design evolved for NASA satellite communications in 2006 \cite{NASA_Paperclip}. The following provides the overview and results of our investigation.

The antenna geometry consists of multiple, unit-length segments connected sequentially. Due to the segmented and bent nature of the resulting antenna, we call this design a ``paperclip antenna.'' Each segment can point toward any direction, thus, the genes that define each individual are the rotation angles between $0$ and $2\pi$ about the three Cartesian axes for each segment. The final individual geometry consists of a number of randomly rotated unit vectors attached tip to tail. The rotations are initially uniformly distributed from $0$ and $2\pi$.

\begin{figure}
    \centering
    \subfloat[]{{\includegraphics[width=.35\linewidth]{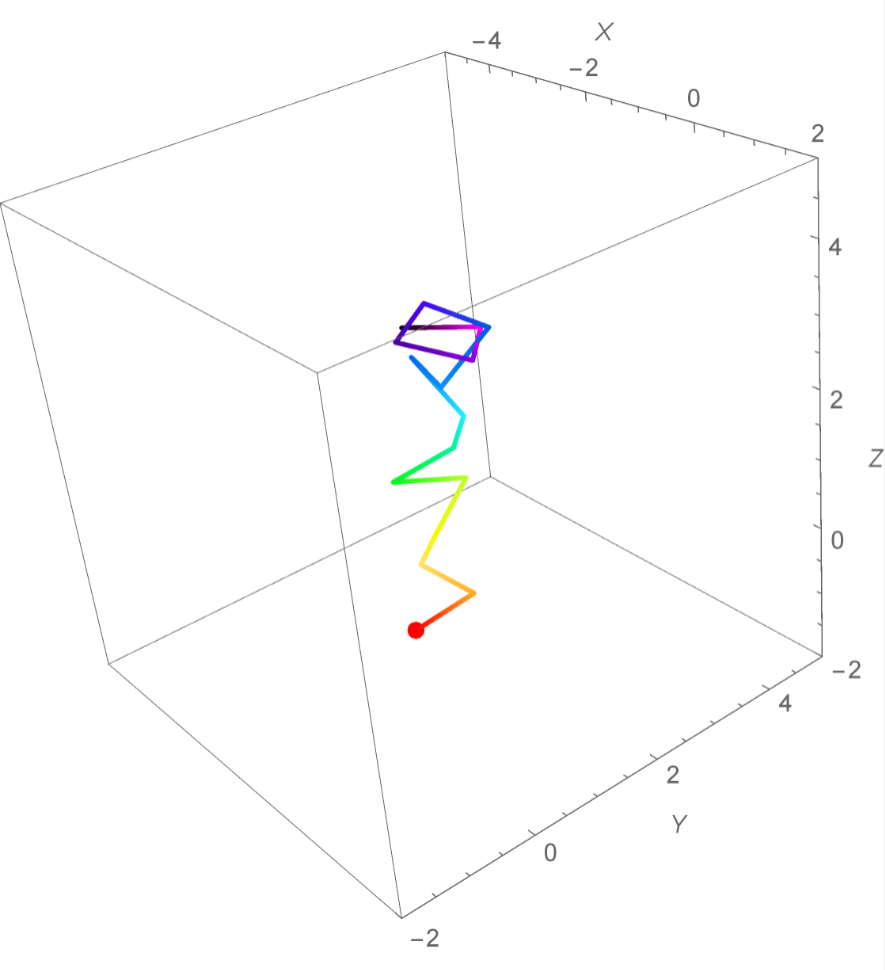} }} 
    \subfloat[]{{\includegraphics[width=.58\linewidth]{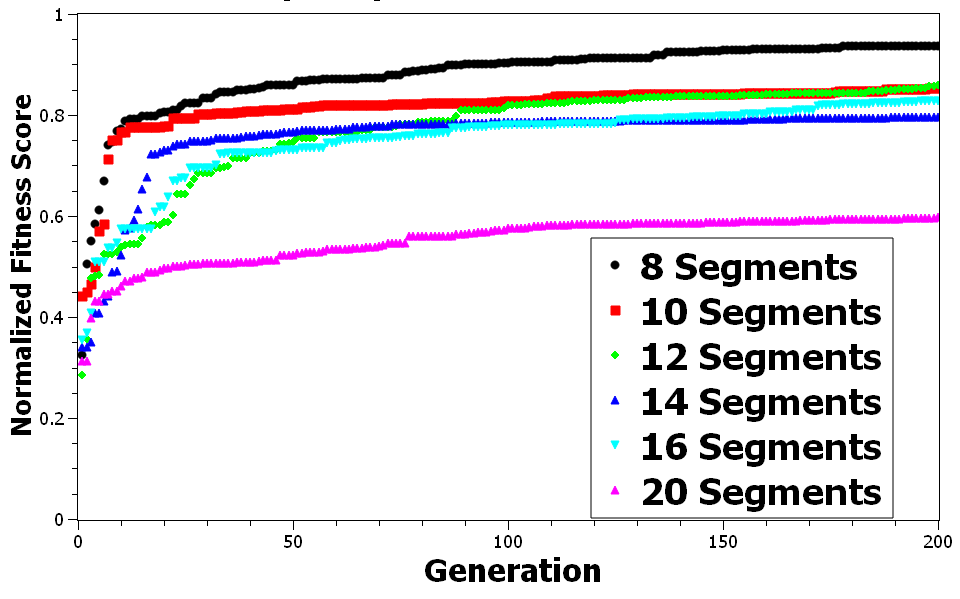} }}%

    \caption{(a) Example of a partially evolved paperclip antenna. Note the general counterclockwise spiral. (b) The best fitness score of 100 paperclip antennas over 200 generations evolved to produce a counterclockwise curl. The GA was performed for various number of segments.   }%
    \label{fig:PaperclipResults}
\end{figure}

While a number of different fitness functions were explored, one in particular directed the evolution of the antenna to arrive to the desired curl shape, as illustrated in Fig. \ref{fig:PaperclipResults}. This ``curl'' function was sensitive to changes in the initial parameters, complex in that all of the evolved rotations had to work together to produce the final shape, and consistent across multiple runs given 200 or fewer generations.

The curl fitness function calculates the cross-product between adjacent vectors, $ \vec{s_i}$ and $\vec{s_{i+1}} $. The fitness score, $F$, is then defined by the equation below:
$$F = \sum_{i=1}^{n-1} \vec{s_i}\times\vec{s_{i+1}}$$
Defined this way, the angle between neighboring antenna segments is preferred to be ninety degrees and oriented counterclockwise in the x-y plane.


The paperclip antennas were evolved over 200 generations composed of 100 individuals using a tournament selection method, with both mutation and crossover operators. This was performed for antennas with various numbers of segments. The results of this analysis are presented in Fig. \ref{fig:PaperclipResults}, which shows the highest scoring designs, per segment, per generation. As shown, fewer segments results in achieving a higher fitness score over fewer generations. A higher quantity of segments increases the complexity of the antennas, thereby resulting in a slower approach to the desired solution.

\section{Antenna Response Evolution Algorithm}
Designing beam patterns by hand can be extremely difficult. AREA is an algorithm developed for the evolution of neutrino detecting, RF antenna responses. There are a number of known, desired properties of antennas, such as sensitivity to specific frequencies, sensitivity to the active volume, and reduced sensitivity to the inactive volume. AREA employs these properties over gain or phase to evolve antenna beam patterns to optimized solutions.


AREA uses a linear sum of 13 azimuthally symmetric, spherical harmonic functions to model the gain or phase pattern of an antenna. Since we are assuming azimuthal symmetry, we only consider spherical harmonics with the magnetic quantum number $m = 0$. This allows the convoluted form of a gain or phase pattern to be described with the 13 coefficients $a_\ell$ (the orbital angular momentum quantum number) of the spherical harmonics.


AREA generates an initial population in which each individual contains a set of randomly generated genes. These individuals are evaluated by the fitness function, to produce a fitness score. The next generation is created through four different selection method and operator combinations. Half of the population is generated through crossover of two parents chosen through roulette selection. One-sixth is generated by crossover of two parents selected from two 6-way tournaments. 
The remaining population is generated through a Gaussian mutation operator of single individuals, with one-sixth of the total population selected through roulette selection and the remaining one-sixth coming from a 6-way tournament. 
The fitness scores of the new population are evaluated. AREA repeats this process until it reaches a specified number of generations~\cite{Harris}.

\begin{table}[h]    \label{tab:AreaSummary}
\begin{center}
\caption{Summary of the selection methods and operators used in the AREA procedure}
\begin{tabular}{ |c|c|c| } 
    \hline
    Fraction    & Selection Method  & Operator \\ 
    \hline
    \hline
    1/2         & Roulette          & Crossover \\ 
    \hline
    1/6         & Tournament        & Crossover \\ 
    \hline
    1/6         & Roulette          & Mutation \\ 
    \hline
    1/6         & Tournament          & Mutation \\ 
    \hline
\end{tabular}

\end{center}
\end{table}


The fitness score is based on simulated neutrino detection rates for various azimuthal angles. AREA uses a simplified version of AraSim. Developed by the ARA collaboration, AraSim is a Monte Carlo neutrino detection simulator which models neutrinos with energies between $E_{\nu} = 10^{17}-10^{21}$~eV and within a 3-5~km radius of the detectors~\cite{AraSim}. The redesigned tool, AraSimLite, simplifies AraSim's fitness function by omitting the simulation's ray-tracing, noise waveforms, signal polarization, and ice modeling, thereby reducing the computational run time\cite{Harris}. 

AREA uses the effective ice volume, the volume of ice that the detector is sensitive to. AraSim and AraSimLite model this effective ice volume as the inherent fitness function. This is given by $V_{eff} = \frac{V_{gen}}{N_{gen}}\Sigma_{i,trig} \omega_i$, where $V_{eff}$ is the effective volume, $V_{gen}$ is AraSimLite's neutrino interaction scan volume, and $N_{gen}$ is the number of neutrino interactions. $\Sigma_{i,trig} \omega$ is the sum of the weights, or probabilities, of neutrino interactions over the range of energies that trigger the detector. Each weight is given by $\omega(E) = \prod_{i = 1}^{N} e^{\frac{l_i}{L_{int,i}(E)}}.$ The fitness score is then the sum of these weights: 

$$F_{AREA} = \sum_{i=1}^{N} \omega_i(E), ~~\frac{g(\theta_i)}{R_i^2} > P_{th}.$$
Here, $\theta_i$ is the angle between the zenith and the vector from the antenna to the event,
$g(\theta_i)$ is the gain at that angle, $R_i$ is the distance to the event. The weights are only counted in the fitness function and fitness score if the ratio of the gain to the square of the distance to the event ($\frac{g(\theta)}{R_i^{2}}$), representing the attenuation of the signal, exceeds a predetermined threshold $P_{th}$. This threshold has been chosen as $150$~m$^{-2}$, the point where the fitness score for an isotropic radiator would reach half of its maximum possible fitness score~\cite{Harris}. 

\begin{figure}
    \centering
    \includegraphics[width=0.9\linewidth]{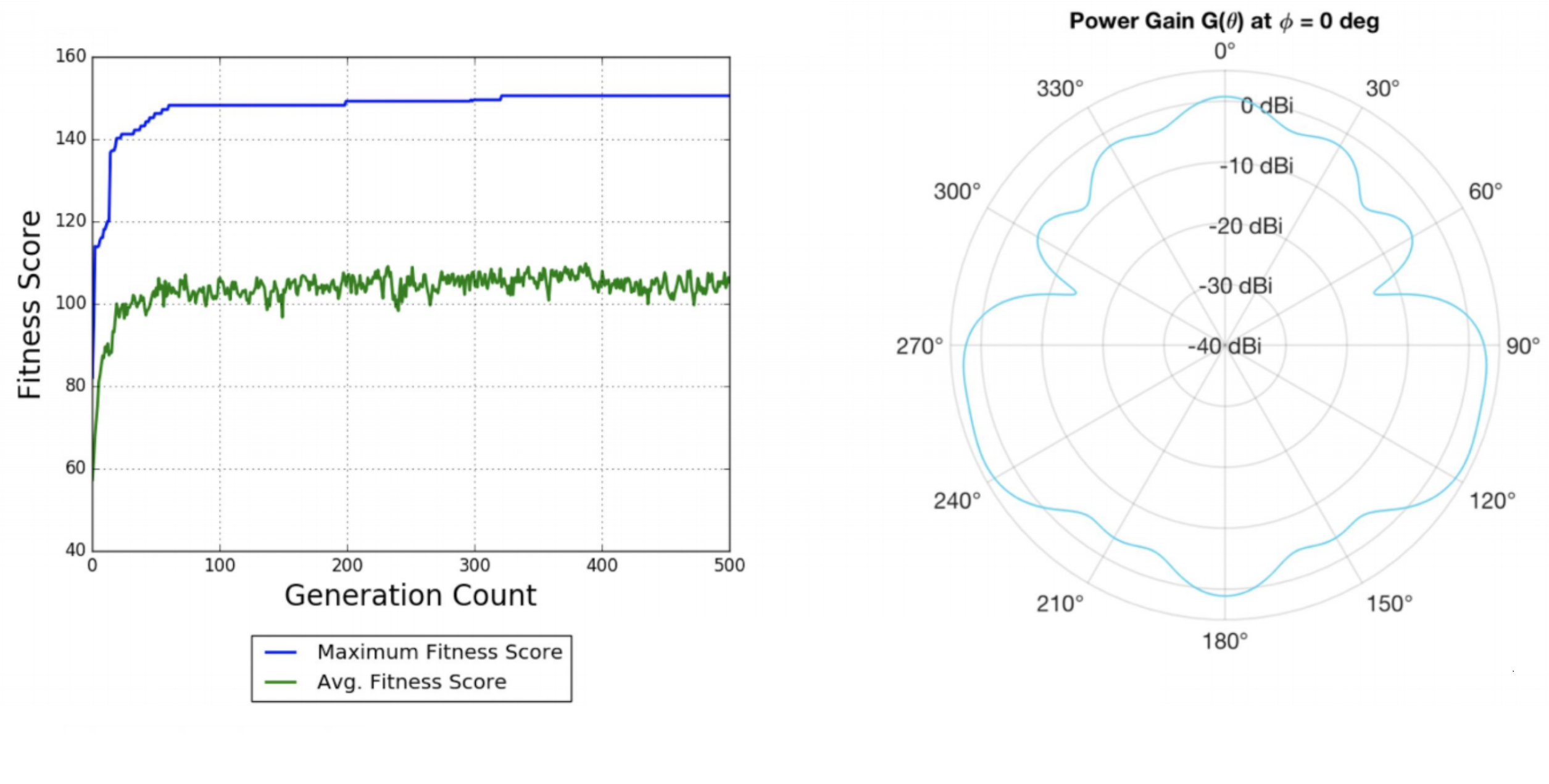}
    \caption{Results of an example AREA procedure for energy $10^{18}$~eV: (a) The change in maximum and average fitness score over 500 generations; (b) The radiation pattern with the best fitness score after 500 generations}
    \label{fig:AreaResults}
\end{figure}

The results of an initial AREA test are presented in Fig. \ref{fig:AreaResults}. On the left, the change in the maximum and average fitness scores of a population are given over 500 generations, showing a convergence in less than 100 generations with minimal improvement thereafter. On the right, a final evolved radiation pattern is shown. The AREA procedure demonstrated the ability to evolve radiation patterns that matched the expected geometry of an antennae designed to detect neutrinos from a particular arrival direction.

\section{Physical Antenna Evolution Algorithm}
The goal of this investigation is to evolve physical antennas to boost neutrino sensitivity. PAEA evolves antenna geometries by means of fitness scores based on the simulation of antenna performance. For physical antenna evolution, a seed generation of possible antenna designs is initialized, with random values picked from a Gaussian distribution. The commercial antenna simulation software, XFdtd is then used to model the associated frequency-dependent response patterns. Those antenna response patterns are input into Monte Carlo neutrino simulation software in order to predict the effective volume of ice the antenna is able to measure. The fitness score for each antenna is fed back into the GA that evolves the geometry of the next generation. This loop continues until a design with acceptable parameters and measurement volume is reached. The PAEA process is generic and can be adapted for many types of antenna designs


\begin{figure}
    \centering
    \includegraphics[width=0.7\linewidth]{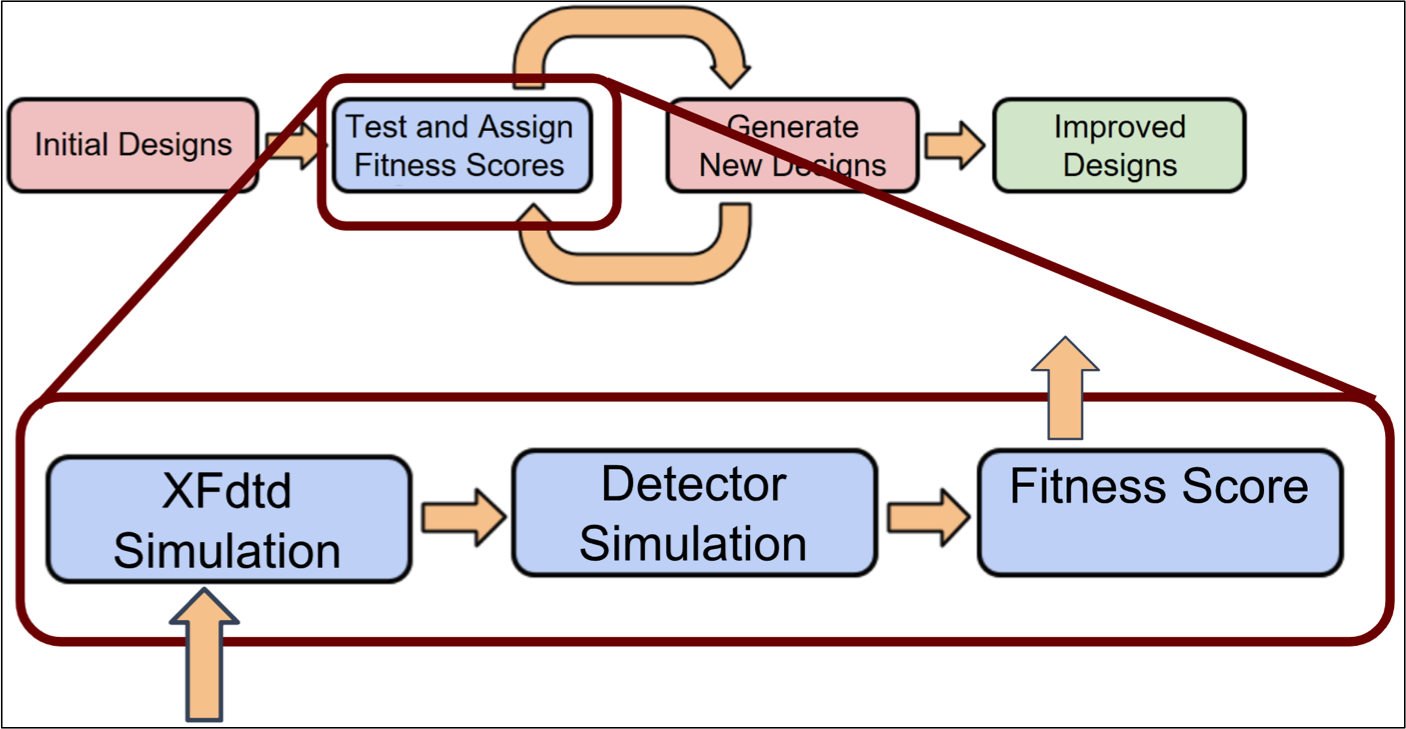}
    \caption{A flow chart of the PAEA procedure with a detailed inset of the fitness score procedure.}
    \label{fig:PAEALoop}
\end{figure}

\begin{figure}
    \centering
    \includegraphics[width=0.25\linewidth]{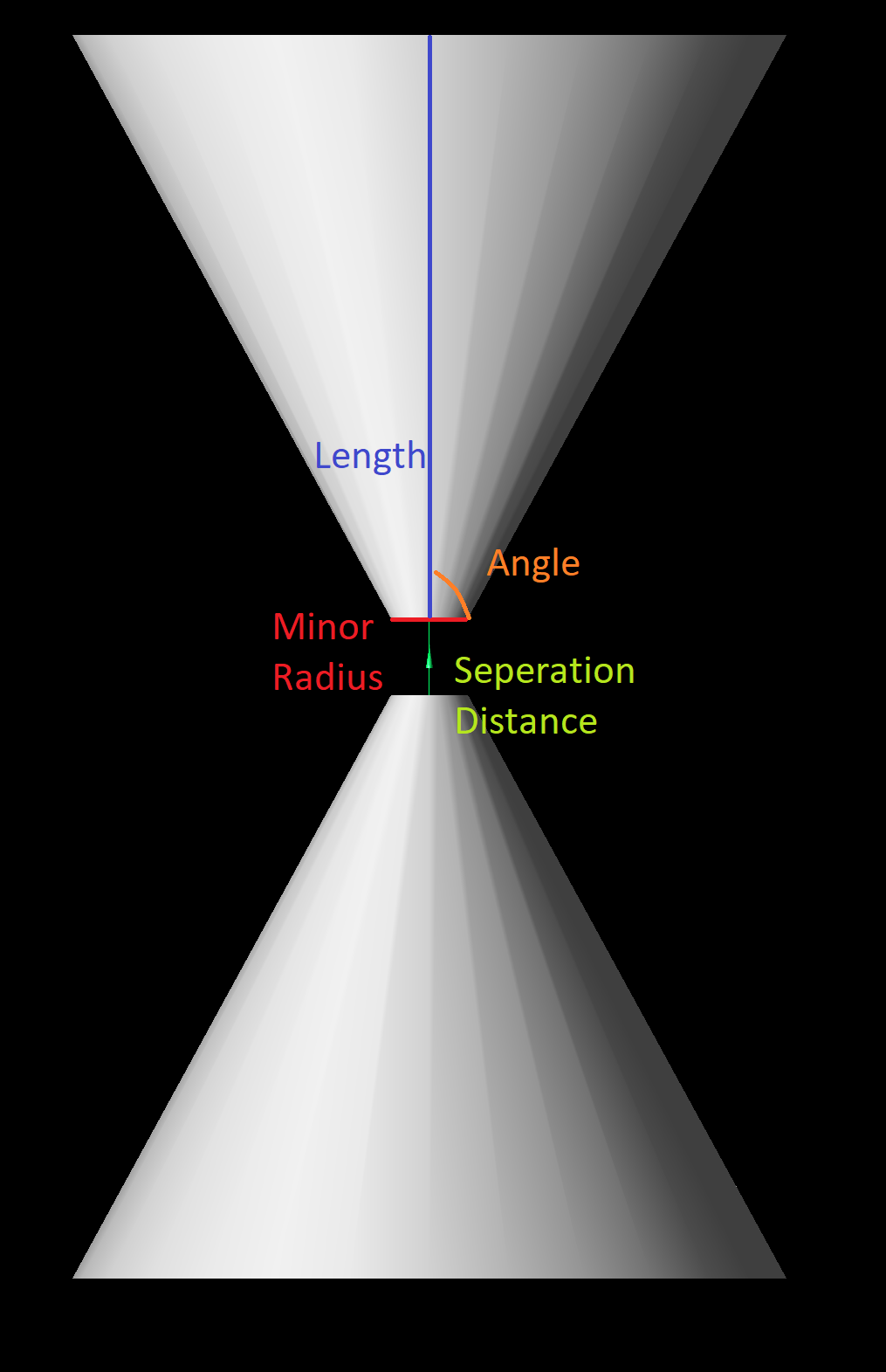}
    \caption{Geometry of bicone antenna showing the genes of length, opening angle and minor radius, and separation distance.}
    \label{fig:BiconeDrawing}
\end{figure}

In the current version of PAEA, we consider a bicone antenna design as shown in Fig.~\ref{fig:BiconeDrawing}. The bicone design consists of two identical truncated cones sitting back-to-back with a fixed separation distance of 2.8~cm. The parameters of the design that are subject to evolution are: the minor radius of the cones, the length of the cones, and their opening angle. 

The evolution of the parameters is conducted through a genetic algorithm
described below that draws from the existing body of work in GAs~\cite{Adaptive_GAs, GA_Practices}. The fitness score attributed to each bicone design is the effective volume of the ARA detector as simulated by AraSim~\cite{Harris, AraSim}.

The GA uses the fitness scores from the individuals in each generation to produce the next generation of individuals using a two step process. The first step uses roulette selection whereby two selected individuals are chosen at random as parents to produce the next generation. Their genes are randomly selected to create each individual offspring. The second step in producing the next generation improves genetic diversity by introducing mutations to $60\%$ of the offspring generated in the first step. This is done by selecting new parameters drawn from a Gaussian distribution whose width is set by the user. The mean and standard deviation of the distribution for each parameter is summarized in Table~\ref{tab:Gaussian}.

\begin{table}    
\begin{center}
\caption{Summary of the mean and standard deviation for the newly selected parameters drawn from a Gaussian distribution}
\label{tab:Gaussian}
\begin{tabular}{ |c|c|c| } 
    \hline
      Parameter & Mean & Standard Deviation \\ 
    \hline
    \hline
    Inner Radius        & 2\,cm          & 0.25\,{\rm cm} \\ 
    \hline
    Length         & 50\,cm        & 15\,{\rm cm}\\ 
    \hline
    Angle         & $\pi$/4          &  $\pi$/6 \\
    \hline
\end{tabular}

\end{center}
\end{table}

\begin{figure}
    \centering
    \subfloat[Generation 0]{{\includegraphics[width=.325\linewidth]{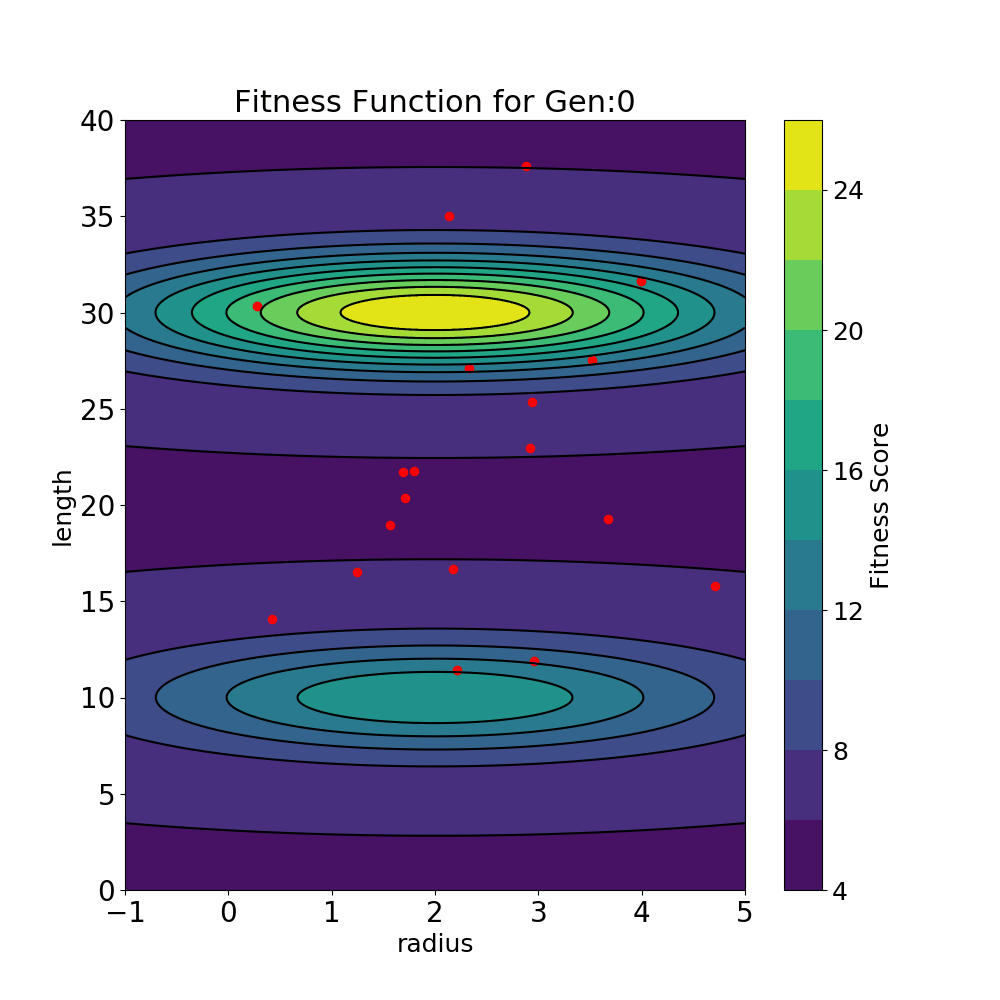} }}%
    \subfloat[Generation 5]{{\includegraphics[width=.325\linewidth]{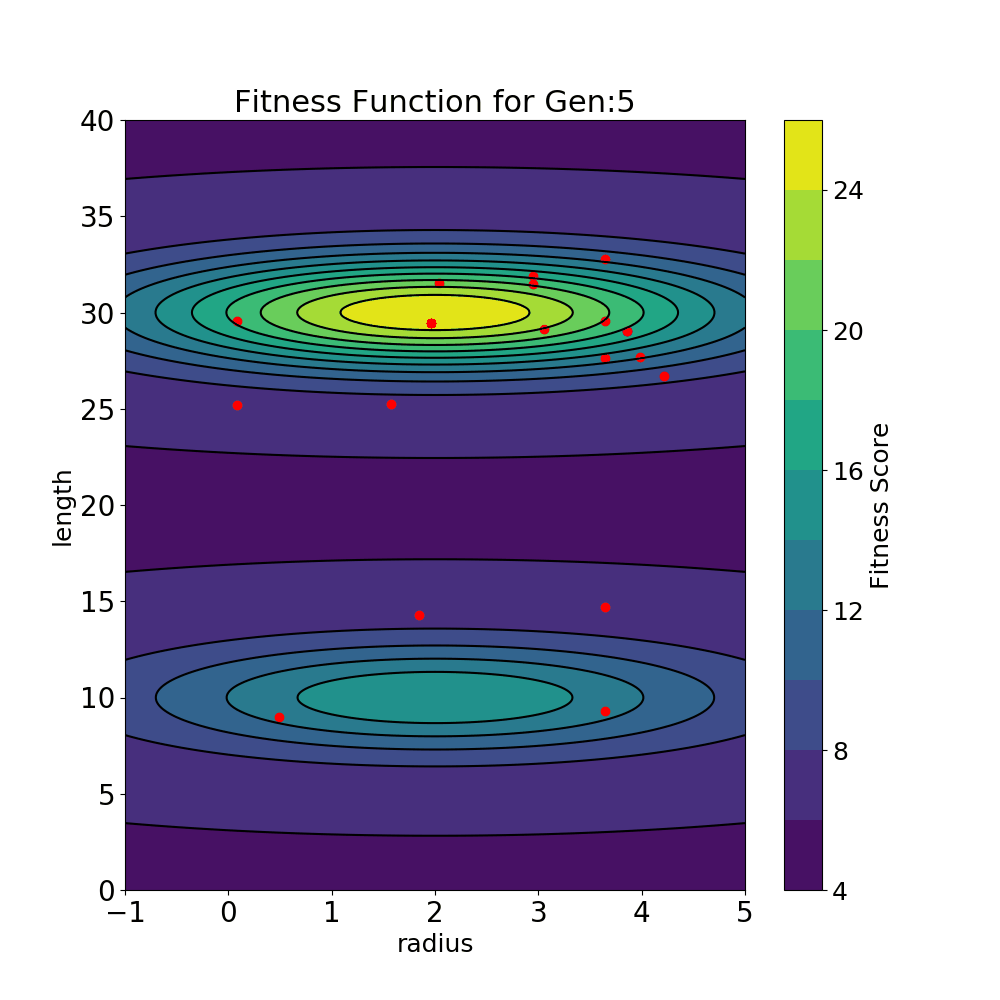} }}%
    \subfloat[Generation 20]{{\includegraphics[width=.325\linewidth]{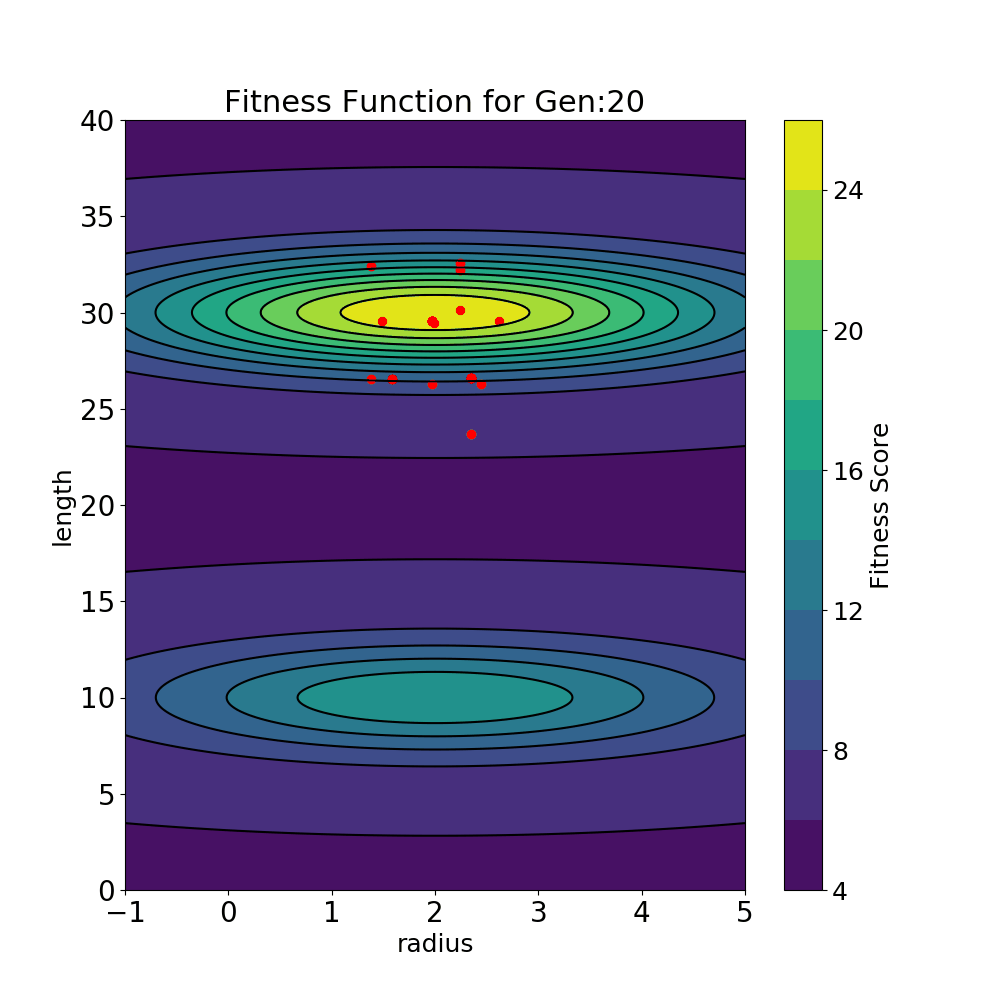} }}%
    \caption{Example of PAEA algorithm results at (a) Generation 0, (b) Generation 5, (c) Generation 20. The fitness score is shown in contour plot. Individuals, shown as red dots, began spread over a wide range, evolved to group near the global maximum. }%
    \label{fig:PaeaResults}%
\end{figure}

Some initial results of PAEA are presented in Fig. \ref{fig:PaeaResults}. In order to test the performance of the algorithm in the presence of local and global maximums, we consider a fitness function that is the sum of two displaced Gaussian distributions of different heights. For this test, we take two parameters that we call length and radius. In the first generation (Generation 0), the 20 individuals, shown as red dots, covered a region of parameter space that contained both Gaussian distributions. By the $20^{th}$ generation, 19 of the 20 individuals were within 2$\sigma$ of the the global maximum, despite some individuals finding the local maximum in earlier generations.

\section{Conclusions}
These proceedings present the initial results of an investigation into the use of GAs to evolve antennas for UHE neutrino detection. Paperclip antennas were successfully evolved into various geometric patterns, one of which was discussed in the body of this research. It is demonstrated that an increase in the complexity of the design, as defined by the quantity of segments, results in a lower, initial fitness score with subsequent, limited improvement in the following generations. Future investigations could include using XFdtd to evolve highly directional paperclip antennas.


The AREA procedure was developed to evolve antenna gain patterns to boost directional sensitivity. A GA was developed to generate optimal gain patterns based on the AraSimLite simulation. Candidate solutions indicate that downward directed antennas should be used in the ARA experiment. Current efforts are focused on using a more advanced neutrino simulation software to construct improved AREA fitness scores and more accurate, resulting gain patterns. 

Finally, the PAEA procedure was used to successfully evolve bicone antenna parameters. Improvements to minimize computation time is a focus of ongoing investigation, thereby enabling higher quality gain patterns for neutrino detection. 
Future research will focus on combining the AREA and PAEA processes for a two-step procedure. AREA would be used to extract optimized antenna response patterns inserted into PAEA to evolve physical antennas. These proceedings demonstrate the potential for GAs to improve antenna beam patterns and antenna geometries, which could be applied to neutrino detectors. Future avenues of research include evolving different types of antennas, array geometries, and trigger systems.

\section*{Acknowledgements}
The GENETIS team is grateful for support from the Ohio State
Department of Physics Summer Undergraduate Research program,
support from the Center for Cosmology and Astroparticle Physics,
and the Cal Poly Connect Grant.
J. Rolla would like to thank the National Science Foundation for 
support under Grant 1404266.  We would also like to thank the Ohio Supercomputing Center.

\bibliographystyle{custom_style}
\bibliography{main}

\end{document}